\documentstyle[12pt]{article}
{\catcode `\@=11 \global\let\AddToReset=\@addtoreset}
\AddToReset{equation}{section}

\makeatletter                                                           

\def\section{\@startsection {section}{1}{\z@}{-1.5ex plus -.5ex         
minus -.2ex}{1ex plus .2ex}{\large\bf}}                                 

\long\def\@makecaption#1#2{\vskip 10pt \setbox\@tempboxa\hbox{#1. #2}   

 \def\@oddfoot{\rm\hfill\thepage\hfill}\def\@evenfoot{\@oddfoot} }      

\newtheorem{Theorem}   {Theorem}   [section]                            
\newtheorem{Corollary} [Theorem]   {Corollary}                          
\newtheorem{Lemma}     [Theorem]   {Lemma}
\newtheorem{Proposition} [Theorem] {Proposition}                        
\newtheorem{Definition} [Theorem] {Definition}

\newcommand{\veps}{{\Sigma}}
\newcommand{\R}{I\!\! R}
\newcommand{\bt}{\begin{Theorem}}
\newcommand{\et}{\end{Theorem}}
\newcommand{\bl}{\begin{Lemma}}
\newcommand{\el}{\end{Lemma}}
\newcommand{\bp}{\begin{Proposition}}
\newcommand{\ep}{\end{Proposition}}
\newcommand{\be}{\begin{equation}}
\newcommand{\ee}{\end{equation}}
\def\bd{\begin{description}}
\def\ed{\end{description}}
\newcommand{\cu}{{\cal U}}

\newcommand{\co}{{\cal{O}}}
\newcommand{\ci}{{\cal{I}}}

\newcommand{\cv}{{\cal{V}}}
\newcommand{\cy}{{\cal{Y}}}
\newcommand{\cw}{{\cal{W}}}
\newcommand{\pr}{{\bf Proof: }}

\newcommand{\sptmetric}{g}
\newcommand{\IR}{I\!\! R}
\newcommand{\eps}{{\epsilon}}
\renewcommand{\part}{{\partial}}
\renewcommand{\a}{{\alpha}}
\newcommand{\Om}{{\Omega}}
\newcommand{\la}{{\lambda}}
\newcommand{\phie}{\emptyset}
\newcommand{\hX}{{\hat X}}
\makeatother                                                            

\title{On completeness of orbits \\
of Killing vector fields}{}{}

\author{
{
P.T.\ Chru\'sciel}\thanks{
On leave  from the Institute of Mathematics of
the Polish Academy of Sciences.
Supported in part by  a KBN
grant \# 2 1047 9101, an NSF grant \# PHY 89--04035 and the Alexander
von Humboldt Foundation. e-mail: piotr@sbitp.ucsb.edu}\\
Institute for Theoretical Physics\\ University of California\\
Santa Barbara, California 93106--4030}

\begin{document}

\maketitle

\begin{abstract}
A Theorem is proved which reduces the problem of completeness of
orbits of Killing vector fields in maximal globally hyperbolic, say vacuum,
space--times
to some properties
of the orbits near the Cauchy surface. In particular it is shown that
all Killing orbits are complete in maximal developements
of asymptotically flat Cauchy data, or of Cauchy data prescribed on a
compact manifold.
\end{abstract}

\section{Introduction}
\label{Section 1}
In any physical theory  a privileged role is played by solutions of
the field equation which exhibit special symmetries. In general
relativity there exist several ways for a solution to be symmetric:
there might  exist
\begin{enumerate}
\item\label{field}
 a Killing vector field $X$ on the space--time
$(M,g)$, or
there might exist
\item\label{globalaction}
 an action of a group $G$ on $M$ by isometries, and finally there might
perhaps exist
\item \label{cauchyaction}
 a Cauchy surface $\Sigma\subset M$ and a group $G$ which
acts on $\Sigma$ while preserving the Cauchy data.
\end{enumerate}It is natural to
enquire what are the relationships between those notions.
Clearly \ref{globalaction}
implies \ref{field}, but  \ref{field} does not need to imply
\ref{globalaction} (remove
{\em e.g.\ }points from a space--time on which an action of $G$
exists).  With a little work one can show
\cite{Moncriefsymmetries,FMM,SCC}
that \ref{cauchyaction} implies \ref{field}, and actually it is true \cite{SCC}
that \ref{cauchyaction} implies \ref{globalaction}, when $M$ is
suitably chosen. The purpose of
this paper is to address the question, {\em do  there exist    natural
conditions on $(M,g)$ under which \ref{field} implies
\ref{globalaction}?} Recall \cite{Ch G} that given  a Cauchy data set
$(\Sigma,\gamma,K)$,
where $\Sigma$ is a three--dimensional manifold, $\gamma$ is a
Riemannian metric on $\Sigma$, and $K$ is a symmetric two--tensor on
$\Sigma$,
there exists a {\em unique up to isometry} vacuum
space--time $(M,\gamma)$,
which is called the {\em maximal globally hyperbolic vacuum development of
$(\Sigma,\gamma,K)$}, with an embedding $i:\Sigma\rightarrow M$ such
that
$i^*g=\gamma$, and such that $K$ corresponds to the extrinsic curvature of
$i(\Sigma)$ in $M$. $(M,\gamma)$ is {\em inextendible} in the class of
globally hyperbolic space--times with a vacuum metric. This class of
space--times is highly satisfactory to work with, as they can be
characterized by their Cauchy data induced on some Cauchy surface.
Let us also recall that in globally hyperbolic vacuum space--times $(M,g)$,
the question of existence of a Killing vector field $X$ on $M$ can be
reduced to
that of existence of appropriate Cauchy data for $X$ on a Cauchy
surface $\Sigma$ ({\em cf.\ e.g.\ }\cite{SCC}). In
this paper we show the following:

\begin{Theorem}
\label{T1}
Let $(M,g)$ be a smooth,  vacuum, maximal globally  hyperbolic space--time with
Killing
vector field $X$ and Cauchy surface $\Sigma$. The following
conditions are equivalent:
\begin{enumerate}
\item
There exists $\epsilon>0$ such that for all $p\in\Sigma$ the orbits
$\phi_s(p)$ of $X$ through $p$ are defined for all
$s\in[-\epsilon,\epsilon]$.
\item
The orbits of $X$ are complete in $M$.
\end{enumerate}
\end{Theorem}

It should be said that though this result seems to be new, it is a
relatively straightforward consequence of the  results in \cite{Ch G}.

The following example\footnote{I am grateful to R.Wald and B.Schmidt
for discussions concerning this point.} shows that some conditions on
the behaviour of the
orbits on the Cauchy surface are necessary  in general: Let $\Sigma$
be a connected component of the unit spacelike hyperboloid in
Minkowski space--time $\IR^{1+3}$,
let $(M,g)$ be
the domain of dependence of $\Sigma$ in  $\IR^{1+3}$ with the
obvious flat metric, let $X$ be the Killing vector $\partial/\partial t$.
$M$ is maximal globally hyperbolic ({\em cf.\ e.g.\,} Proposition
\ref{onmaximality} below), $\Sigma$ is a complete Riemannian manifold, the
Lorentzian length of $X$ is uniformly bounded on $\Sigma$,
nevertheless no orbits of $X$ are complete in $M$.

As a Corollary of Theorem \ref{T1} one obtains, nevertheless, the
following ({\em cf.\,} Section \ref{ProofsC1} for precise definitions):

\begin{Corollary}
\label{C1}
Let $(M,g)$ be a smooth, vacuum, maximal globally hyperbolic space--time with
Killing vector $X$ and with an
achronal spacelike hypersurface $\Sigma$.
Suppose that either
\begin{enumerate}
\item
$\Sigma$ is compact, or
\item
$(\Sigma,\gamma,K)$ is asymptotically flat, 
or
\item
$(\Sigma,\gamma,K)$ are Cauchy data for an asymptotically flat
exterior region in a (non--degenerate) black--hole space--time.
\end{enumerate}
Then the orbits of $X$ are complete in $D(\Sigma)$.
 \end{Corollary}

The difference between cases 2 and 3 above is, roughly speaking, the
following: In point 2 above $\Sigma$ is a complete Riemannian
submanifold of $M$ {\em without boundary}. On the other hand,  in point 3 above
$\Sigma$ is a complete Riemannian
submanifold of $M$ {\em with a compact  boundary} $\partial \Sigma$, and
 the Killing vector is assumed to be tangent to $\partial
\Sigma$; {\em cf.\/} the beginning of Section \ref{ProofsC1} for a longer
discussion of the relevant notions. [It should also be pointed out,
that we necessarily have $M=D(\Sigma)$ in point 1 above by
\cite{BILY}. In point 3 above, however, $M=D(\Sigma)$ {\em cannot}
hold, {\em cf.\/} Definition D2, Section \ref{ProofsC1}.]

We have stated Theorem \ref{T1} and Corollary \ref{C1} in the vacuum,
but they  clearly hold for any kind of well posed hyperbolic system of
equations for the metric $g$ coupled with some matter fields. All that is
needed is a local existence  theorem for the coupled
system, together with  uniqueness of solutions in domains of
dependence. In particular, Theorem \ref{T1}  will still be true {\em
e.g.\/} for metrics satisfying the Einstein -- Yang--Mills -- Higgs 
equations. Corollary \ref{C1} will still hold for  the Einstein --
Yang--Mills -- Higgs equations,
provided 
both the gravitational field and the matter fields satisfy appropriate
fall--off conditions in the asymptotically flat case.
We plan to discuss this elsewhere.



The use and applicability of Theorem \ref{T1} and Corollary \ref{C1}
is rather wide:   
all non--purely--local results
about
space--times with Killing vectors  assume completeness of their orbits.
Let us in particular mention the  theory of uniqueness of
black--holes. 
Clearly it is essential to also classify those black--holes
in which the orbits of the Killing field are not
complete, and which are thus not covered by the existing theory.
Consider then a stationary black hole space--time $(M,g)$ in  which an
asymptotically flat
Cauchy surface exists
but in which the Killing orbits are {\em not}
complete:   Corollary
\ref{C1} shows that $(M,g)$ can be enlarged to
obtain  a space--time with complete Killing orbits.
As another application, let us also mention the recent work of Wald
and this author \cite{ChW}, where the question of existence of maximal
hypersurface in
asymptotically stationary space--times is considered. Corollary
\ref{C1} shows that the hypothesis of completeness of the orbits of
the Killing field made in \cite{ChW} can be removed, when the
space--time under consideration is vacuum (or  satisfies some well
behaved field equations) and {\em e.g.\ }maximal.

%
%
{\bf Acknowledgements.} Most of the work on this paper was done when
the author was visiting the Max Planck Institut f\"ur Astrophysik in
Garching; he is grateful to J\"urgen Ehlers and to the members of the
Garching relativity group for hospitality. Useful discussions with
Berndt Schmidt and Robert Wald are acknowledged.

\section{Proof of Theorem {\protect\ref{T1}}}
\label{Proofs}
In this Section we shall prove  Theorem \ref{T1}. Let us start with a
somewhat weaker result:

\begin{Theorem}\label{T1.0}
 Let $(M,  \sptmetric$) be a smooth, vacuum, maximal globally
hyperbolic space-time with Cauchy surface $\Sigma$ and
with a Killing vector field X, with $\sptmetric, X \in
C^\infty$.  Then the orbits of $X$ in $M$ are complete
if and only if
\bd\item  [(i)] there exists $\epsilon > 0$ such that for
all $p \in \Sigma$ the orbits $\phi_s(p)$ of $X$
are defined for all $s \in [ - \epsilon, \epsilon]$, and
\item [(ii)] for $s \in [-\epsilon,
\epsilon]$ the sets $\phi_s(\Sigma)$ are achronal.
\ed
\et

{\bf Proof:}  Let us start by showing necessity: Point (i)
is obvious, consider point (ii).  As the orbits of $X$
are complete, the flow of $X$ (defined as the solution
of the equations ${d\phi_s(p)\over ds} = X\circ \phi_s(p)$,
with initial value $\phi_0 (p) = p$) is defined for all $p
\in M$ and all $s\in \IR$.  Suppose there exists
$s_1 \in M$ and a timelike path $\Gamma : [0,1]
\rightarrow M$ with $\Gamma(0) \in \phi_{s_1} ( \Sigma)$ and
$\Gamma(1) \in \phi_{s_1} (\Sigma)$. Then
$\phi_{-s_1}(\Gamma)$ would be a timelike path with
 $\phi_{-s_1}(\Gamma(0))\in \Sigma, \phi_{-s_1}(\Gamma (1)) \in
\Sigma $, which is not possible as $\Sigma$ is achronal.
Hence (i) and (ii) are necessary.

To show sufficiency, we shall need the following proposition:

\bp\label{P1}
Let $(M_a, \sptmetric_a) $, $a = 1, 2$, be vacuum globally hyperbolic
space-times with
Cauchy surfaces $\Sigma_a$, and suppose that ($M_2, \sptmetric_2$) is maximal.
Let $\co \subset M_1$ be a (connected) neighbourhood of
$\Sigma_1$ and
suppose there exists a one-to-one isometry $\Psi_{\co} : \co
\rightarrow M_2$,
such that $\Psi_{\co}(\Sigma_1)$ is achronal.  Then there exists a
one-to-one isometry
\be
\Psi : M_1 \rightarrow M_2,
\ee
such that $\Psi |_{\co} = \Psi_{\co}$.
\ep
 {\bf Remarks:}
\begin{enumerate}
\item
 When $\Psi_{\co}(\Sigma_1) = \Sigma_2$, this result can be
essentially found
in \cite{Ch G}.  The proof below is a rather straightforward
generalization of  the
arguments of \cite{Ch G}, {\em cf.\ }also \cite{Ch Y,HE}. Although we assume
smoothness of the metric throughout this paper for the sake of
simplicity, we have taken some care to write the
proof below in a way which
generalizes with no essential difficulties to the case where 
low Sobolev--type differentiability of the metric is assumed.
\item
The condition that $\psi_{{\cal O}}(\Sigma_1)$ is achronal is
necessary, which can be seen as follows:
Let $M_1=\R^2$ with the standard flat metric, set $\Sigma_1=\{t=0\}$. Let
$\sim_a$ be the equivalence relation defined as $(t,x)\sim_a(t+a,x+1)$,
where $a$ is a number satisfying $|a|<1,\ a\neq0$. Define
$M_2=M_1/{\sim_a}$ with
the naturally induced metric, ${\cal O}=(-a/3,a/3)\times\R$, $\psi_{{\cal
O}}=i_{M_1}|_ {{\cal O}}$, where
$i_{M_1}$ is the natural projection: $i_{M_1}(p)=[p]_{\sim_a}$. $M_2$ is
causal geodesically complete; the function $t-ax:M_1\to \R$ defines,
by passing to
the quotient, a time function on $M_2$ the level sets of which are Cauchy
surfaces. It follows that $M_2$ is maximal globally hyperbolic. Clearly
$\psi_{{\cal O}}(\Sigma_1)$ is not achronal, and there is no one-to-one
isometry
from $M_1$ to $M_2$.
\end{enumerate}

\pr Consider the collection $ {{\cal X}}$ of all pairs $(\cu,
\Psi_\cu)$, where
$\cu \subset M_1$ is a globally hyperbolic neighbourhood of $\Sigma_1$ (with
$\Sigma_1$
 -- Cauchy surface for $(\cu,  \sptmetric_{1}|_{\cu})$, and
$\Psi_{\cu}:\cu\rightarrow M_2$ is an isometric
diffeomorphism between $\cu$ and $\Psi_{\cu}(\cu)\subset M_2$ satisfying
$\Psi_{\cu}|_{\Sigma_1} = \Psi_{\co}|_{\Sigma_1}$.  ${\cal{X}}$ can be
ordered by
inclusion: $(\cu,
\Psi_{\cu}) \leq (\cv, \Psi_{\cv})$ if  $\cu \subset \cv$ and if
$\Psi_{\cv}|_{\cu} =
\Psi_{\cu}$.  Let $(\cu_\alpha, \Psi_\alpha)_{\alpha \in\Omega}$ be a
chain in $ {\cal{X}}$,
set $\cw = \cup_{\alpha\in\Omega} \cu_\alpha$, define $\Psi_{\cw} :
\cw \rightarrow
M_2$ by  $\Psi_{\cw}|_{\cu_\alpha} = \Psi_\alpha$;   clearly ($\cw,
\Psi_{\cw}$) is a
majorant for $(\cu_\alpha, \Psi_\alpha)_{\alpha \in\Omega}$. From the
set theory
axioms ({\em cf.\ e.g.\/} \cite{Kelley}[Appendix]) it is easily seen
that $ {{\cal X}}$ forms
a set, we can thus apply Zorn's  Lemma \cite{Kelley} to conclude that
there exist maximal elements
$(\tilde{M}, \Psi)$ in $ {{\cal X}}$. Let then $(\tilde{M}, \Psi)$ be
any maximal element, by definition 
($\tilde{M},
\sptmetric_{1}|_{\tilde{M}}$) is thus globally hyperbolic with Cauchy
surface $\Sigma_1$,
and $\Psi$ is a one-to-one isometry from $\tilde{M}$ into $M_2$ such that
$\Psi|_{\Sigma_1} = \Psi_{\co}|_{\Sigma_1}$.  By {\em e.g.\ }Lemma 2.1.1
of \cite{SCC} we have
\be
 \Psi |_{\tilde{M}\cap \co}  = \Psi_{\co}|_{\tilde{M}\cap\co} . \label{(1)}
\ee
We have the following:
\bl
\label{L1}
 Under the hypotheses of Proposition \ref{P1}, suppose that $(\co,
\Psi_{\co})$  is
maximal.  Then the manifold
$$M^\prime = (M_1 \sqcup M_2)/\Psi_{\co}$$
is Hausdorff.
\el

{\bf  Remark: }
  Recall that $\sqcup$ denotes the disjoint union, while $(M_1\sqcup
M_2)/\Psi$ is the   quotient manifold $(M_1 \sqcup M_2 )/  \sim, $ where
$p_1 \in M_1$
is equivalent to $p_2 \in M_2$ if $p_2 = \Psi(p_1)$.

{\bf Proof:}  Let $p, q \in M^\prime$ be such that there exist no open
neighbourhoods
 separating $p$ and $q$; clearly this is possible only if
(interchanging $p$ with
$q$ if necessary) we have $p\in \partial \co$ and $q\in \partial
\Psi_{\co}({\co})$.
Consider the set ${\cal H}$ of   ``non-Hausdorff points" $p^\prime$ in
$M^\prime$ such that $p^\prime = i_{M_1}(p)$  for some $p \in M_1$,
where $i_{M_1}$ is
the    embedding of $M_1$ into $M^\prime$; ${\cal H}$ is closed and
we have ${\cal H} \subset \partial \co$.

Suppose that  ${\cal H}   \not= \emptyset$, changing time orientation
if necessary we
may assume that ${\cal H}\cap
I^+(\Sigma_1)\ne\emptyset$; let  $p^\prime \in {\cal H} \cap
I^+(\Sigma_1)$.  We wish to show
that there necessarily exists $p \in  {\cal H}$  such that
\be
 {J}^-(p) \cap  {\cal H}\cap I^+(\Sigma_1) = \{ p\}. \label{(2)}
\ee
If (\ref{(2)}) holds with $p = p^\prime$ we are done, otherwise consider the
(non-empty) set
$\cy$ of causal paths $\Gamma: [0,1] \rightarrow I^+ (\Sigma)$ such that
$\Gamma (0) \in {\cal H}, \Gamma(1) = p^\prime$.  $\cy$ is directed by
inclusion:
$\Gamma_1 < \Gamma_2$ if $\Gamma_1  ([0,1])  \subset \Gamma_2 ([0,1]) $. Let
$\{\Gamma_\alpha\}_{\alpha \in \Omega}$ be a chain in ${\cal Y}$, set
$\Gamma =
\cup_{\alpha\in \Omega} \Gamma_\alpha ([0,1]) $, consider the
sequence $p_\alpha =
\Gamma_\alpha(0)$.  Clearly $\Gamma\subset J^+(\Sigma_1) =
I^+(\Sigma_1) \cup \Sigma_1$,
and global hyperbolicity implies that
$\Gamma$ must be extendible, thus $\Gamma_\alpha(0)$ accumulates at
some $p_* \in
I^+(\Sigma_1)\cup \Sigma_1$. As
 $ \co$ is an open neighbourhood of $\Sigma_1$ the case $p_*  \in
\Sigma_1$ is not
possible, hence $p_*  \in I^+(\Sigma_1)$ and consequently $\Gamma \in \cy$.
It follows that every chain in ${\cal Y}$ has a majorant, and by
Zorn's Lemma ${\cal Y}$
has maximal elements.  Let then $\Gamma$ be any maximal element of
${\cal Y}$, setting
$p = \Gamma(0)$ the equality (\ref{(2)}) must hold.

We now claim that (\ref{(2)}) also implies
\be
{J}^-(p) \cap \partial \co \cap I^+(\Sigma_1)= \{ p\}. \label{(3)}
\ee
Suppose, on the contrary, that there exists $q \in (J^-(p)\cap
\partial\co\cap I^+(\Sigma_1))\setminus
\{p\}$; let $q_i \in \co$ be a sequence such that $q_i \rightarrow q $. We
can choose $q_i$ so that $q_{i+1} \in I^+(q_i)$. Global hyperbolicity
of $\co$ implies
that for $i >  i_o$, for some $i_o$, there exist timelike paths
$\Gamma_i : [0,1]
\rightarrow \bar{\co}$, $\Gamma_i \Big([0,1)\Big)\subset \co$, $\Gamma_i(0)  =
q_i $, $\Gamma_i(1)  =
p $. Let $\tilde p\in M_2$ be a non--Hausdorff partner of $p$ such that
the curves $\Psi_\co\Big(\Gamma_i\Big([0,1)\Big)\Big)$ have $\tilde p$
as an accumulation point. 
We have $\Psi_\co(\Gamma_i)\subset J^+\Big(\Psi_\co(q_{i_0})\Big)\cap
J^-\Big(\tilde p)$ 
which is compact by global hyperbolicity of $M_2$,
hence there exists a subsequence $\Psi_\co(q_i)$ converging to some
$\tilde q\in M_2$.
This implies that $q$ and $\tilde q$
constitute a ``non-Hausdorff pair" in $M^\prime$, contradicting
(\ref{(2)}), and thus
(\ref{(3)}) must be true.

Let $p_1\in M_1$, $p_2\in M_2$, $i_{M_{1}}(p_1) = i_{M_{2}}(p_2)$, be
any
non-Hausdorff pair in $M^\prime$ such that (\ref{(2)}) holds with $p =
p_1$.  Let
$x^\mu$ be harmonic coordinates defined in a neighbourhood $\co_2$ of
$p_2$.  [Such
coordinates can be {\em e.g.\ }constructed as follows: Let $t_0$ be
any time
function defined in some neighbourhood $\co_2$ of $p_2$, such that
$t_0(p_2)=0$, set $\ci_\tau
= \{p\in\co_2
: t_0(p) = \tau\}$.  Passing to a subset of $\co_2$ if necessary there exists a
global coordinate system $x^i_0$ defined on $\ci_{0}$;
again passing
to a subset of $\co_2$ if necessary we may assume that $\co_2$ is
globally hyperbolic
with Cauchy surface $\ci_{0}$.
Let $x^\mu\in C^\infty(\co_2)$ be the (unique) solutions of the problem
$$
\Box_{g_{2}} x^\mu = 0,
$$
\be
x^0\bigg|_{\ci_{0}} = 0, \quad {\partial x^0\over\partial
t}\bigg|_{\ci_0} =
1, \quad
x^i\bigg|_{\ci_0} = x^i_0, \quad
{\partial x^i\over \partial t}\bigg|_{\ci_0} = 0,
\ee
where $\Box_\gamma$ is the d'Alembert operator of a metric $\gamma$.
Passing once
more to a globally hyperbolic
 subset of $\co_2$ if necessary, the functions $x^\mu $ form a coordinate
system on $\co_2$.]  We can choose $\epsilon > 0$ such that
\begin{enumerate}
\item $\mbox{int} 
D^+(\ci_{
-\epsilon})
\subset
\co_2$,
\item $p \in \mbox{int} 
D^+(\ci_{
-\epsilon})
$,
\item $\overline{\ci_{
-\epsilon}} \subset \Psi_\co(\co). $
\end{enumerate}
Define
$${\hat \ci}  = \Psi^{-1}_\co (\ci_{
-\epsilon}).
$$
Let $y^\mu  \in C^\infty \Big(D({\hat \ci} )\Big)$ be the (unique)
solutions of the problem
$$
\Box_{g_{1}}y^\mu = 0, $$
$$
y^\mu\bigg|_{\hat \ci}  = x^\mu\circ  \Psi_\co\bigg|_{{\hat\ci} },
\quad {\partial y^\mu\over \partial
 {\hat n} }= {\partial \left(x^\mu \circ\Psi_\co\right)\over \partial
\hat n
 } \Big|_{\hat\ci},$$
where ${\partial\over\partial {\hat n} }$ is the derivitive in the direction
normal to
${\hat \ci} $. By isometry invariance of the wave equation we have
\be
y^\mu|_{D({\hat\ci} )\cap \co} = x^\mu \circ \Psi_\co|_{D(
{\hat \ci} )\cap\co}.
\label{(NEQ.1)}
\ee
Set
$$\cu = \co \cup \mbox{int} D^+({\hat \ci} ), $$
and for $p\in\cu$ define
\be
\Psi_\cu(p) = \cases{\Psi_\co(p), & $p\in\co$,\cr
  q: \mbox{where $q$ is such that\ } x^\mu(q) = y^\mu (p), &$p\in
\mbox{int} D^+({\hat \ci} )$.\cr}  \ee
{}From (\ref{(NEQ.1)}) it follows that $\Psi_\cu$ is a smooth map from
$\cu$ to $M_2$.
Clearly $\cu$ is a globally hyperbolic neighbourhood of
$\Sigma_1$, and $\Sigma_1$
is a Cauchy surface for $\cu$.  Note that $\co$ is a proper subset of
\/$\cu$, as
$p_1\in \mbox{int} D^+({\hat \ci} )$ but $p_1 \not\in \co$. It follows
from
uniqueness of solutions
of Einstein equations in harmonic coordinates that $\Psi_\cu$ is an
isometry.  To prove
that $\Psi_\cu$ is one-to-one, consider $p, q \in \cu$ such that
$\Psi_\cu(p) =
\Psi_\cu(q)$. Changing time orientation if necessary we may suppose
that $p \in
I^+(\Sigma_1)$.  By hypothesis we have $I^+\Big(\Psi_\co(\Sigma_1)\Big)\cap
I^-\Big(\Psi_\co(\Sigma_1)\Big)
= \emptyset$, hence $q \in I^+(\Sigma_1)$.
Let $[0,1] \ni s \rightarrow \Gamma(s)$ be a timelike path from
$\Sigma_1$ to
$q$, let $\Gamma_1(s)$ be a connected component of
$\Psi^{-1}_\cu(\Psi_\cu(\Gamma))$
which contains $\{p\}$. Consider the set $\Omega = \{ s\in [0,1] :
\Gamma(s)=\Gamma_1(s)\}$.
Since $\Psi_\cu|_\co = \Psi_\co$ which is one-to-one, $\Omega$ is
non-empty.  By
continuity of $\Gamma_1$ and $\Gamma$, $\Omega$ is closed.  Since
$\Psi_\cu$ is
locally one-to-one (being a local diffeomorphism), $\Omega$ is open.
It follows that
$\Omega = [0,1]$, hence $p = q$, and $\Psi_\cu$ is one-to-one as
claimed.

We have thus shown, that $(\co, \Psi_\co) \leq (\cu, \Psi_\cu)$
and $(\co, \Psi_\co) \neq (\cu, \Psi_\cu)$ which contradicts
maximality  of $(\co, \Psi_\co)$.  It follows that
$M^\prime$ is Hausdorff, as we desired to show.   $\hfill\Box$


%
%
%
%
%

Returning to the proof of Proposition~\ref{P1}, let ($\tilde M,\Psi$)
be maximal. If
$\tilde M=M_1$ we are done, suppose then that $\tilde M\neq M_1$. Consider the
manifold
$$
M'=(M_1\sqcup M_2)/\Psi  .
$$
By Lemma \ref{L1}, $M'$ is Hausdorff. We claim that $M'$ is globally hyperbolic
with Cauchy surface $\veps'=i_{M_2}(\veps_2)\approx \veps_2$, where
$i_{M_a}$ denotes the  canonical embedding of $M_a$ in $M'$.
Indeed, let $\Gamma'\subset M'$ be an inextendible causal curve in $M'$, set
$\Gamma_1=i_{M_1}^{-1}(\Gamma'\cap i_{M_1}(M_1))$,
$\Gamma_2=i_{M_2}^{-1}(\Gamma'\cap i_{M_2}(M_2))$. Clearly
$\Gamma_1\cup\Gamma_2\neq\phie$, so that either $\Gamma_1\neq\phie$, or
$\Gamma_2\neq\phie$, or both. Let  the index $a$ be such that
$\Gamma_a\neq\phie$.
If $\hat\Gamma_a$ were an extension of $\Gamma_a$ in $M_a$, then
$i_{M_a}(\hat\Gamma_a)$ would be an extension of $\Gamma'$ in $M'$, which
contradicts maximality of $\Gamma'$, thus $\Gamma_a$ is inextendible. Suppose
that $\Gamma_1\neq\phie$; as $\Gamma_1$ is inextendible in $M_1$ we must have
$\Gamma_1\cap\veps_1=\{p_1\}$ for some $p_1\in \veps_1$.
We thus have
$\Psi(p_1)\in \Gamma_2$, so that 
it always holds that $\Gamma_2\neq\phie$.
By global hyperbolicity of $ M_2$ and inextendibility of $\Gamma_2$ it follows
that $\Gamma_2\cap\veps_2=\{p_2\}$ for some $p_2\in \veps_2$, hence
$\Gamma'\cap i_{M_2}(\veps_2)=\{i_{M_2}(p_2)\}$. This shows that
$i_{M_2}(\veps_2)$
is a Cauchy surface for $M'$, thus $M'$ is globally hyperbolic. As $\tilde
M\neq M_1$ we have $M'\neq M_2$ which contradicts maximality of $M_2$. It
follows that we must have $\tilde M=M_2$, and Proposition~\ref{P1} follows.
\hfill$\Box$

Returning to the proof of Theorem \ref{T1.0}, choose $s\in [-\eps/2,\eps/2]$.
There
exists a globally hyperbolic neighborhood ${\cal O}_s$ of $\veps$ such
that the map
$\phi_s(p)$ is defined for all $p\in  {\cal O}_s$:
$$
{\cal O}_s\ni p\to\phi_s(p)\in  M  .
$$
$\phi_s(\veps)$ is achronal by hypothesis, and Proposition~\ref{P1} shows
that there exists
a map $\hat\phi_s:M\to M$ such that $\hat\phi_s|_{{\cal O}_s}=\phi_s$.
For $s\in  I\!\!R$ let
$k$ be the integer part of $2s/\eps$, define $\hat\phi_s:M\to M$ by
$$
\hat\phi_s=\hat\phi_{s-k\eps/2}\circ
\underbrace{\hat\phi_{\eps/2}\circ\dots\circ\hat\phi_{\eps/2}}_{k{\rm\
times}} .
$$
It is elementary to show that $\hat\phi_s$ satisfies
$$
{d\hat\phi_s\over ds}=X\circ\hat\phi_s,
$$
and Theorem \ref{T1.0} follows. \hfill $\Box$

Let us point out the following useful result:

\begin{Proposition}\label{P2}
\label{onmaximality}
Let $(M,g)$ be a maximal globally hyperbolic vacuum space-time.
Suppose that $\tilde\veps \subset M$ is an achronal spacelike
submanifold, and let
$(\tilde\gamma,\tilde K)$ be the Cauchy data induced by $g$ on $\tilde\veps$.
Then $(D(\tilde\veps),g|_{D(\tilde\veps)})$ is isometrically
diffeomorphic to the
maximal globally hyperbolic vacuum development $(\tilde M,\tilde\gamma)$ of
$(\tilde\veps,\tilde \gamma,\tilde K)$.
\end{Proposition}

{\bf Proof:} 
By maximality of  $(\tilde M,\tilde\gamma)$, there exists a map $\Psi: {\cal
D}(\tilde\veps)\to \tilde M$ which is a smooth isometric diffeomorphism between
$D(\tilde\veps)$ and $\Psi(D(\tilde\Sigma))$. By standard local
uniqueness results for
vacuum Einstein equations there exists a globally hyperbolic neighborhood
$\tilde {\cal O}$ of $\tilde\imath(\tilde\veps)$ in $\tilde M$, where
$\tilde\imath$
is the embedding of $\tilde\veps$ in $\tilde M$, and a map
$\Phi_{\tilde {\cal O}}:\tilde {\cal O}\to{\cal
D}(\tilde\veps)\subset M$ which is an isometric
diffeomorphism between $\tilde {\cal O}$ and $\Phi_{\tilde {\cal O}}(\tilde
{\cal O})$.
By Proposition~\ref{P1}, $\Phi_{\tilde {\cal O}}$ can be extended to a map
$\Phi:\tilde
M\to M$ which is an isometric diffeomorphism between $\tilde M$ and
$\Phi(\tilde
M)$.
Clearly we must have $\Phi(\tilde M)\subset D(\tilde\veps)$, so that
one obtains
$\Psi \circ\Phi=id_{\tilde M}$, $\Phi\circ\Psi=id_{D(\tilde\veps)}$,
and the result
follows. \hfill $\Box$

To Prove Theorem~\ref{T1}, {\it i.e.,\/} to remove the hypothesis (ii) of
Theorem~\ref{T1.0}, more work is needed. Let
$t_\pm(p)\in  \R \cup\{\pm\infty\}$, $t_-(p)<0<t_+(p)$ be defined by the
requirement that $(t_-(p),t_+(p))$ is the largest connected interval
containing 0 such that the solution $\phi_s(p)$ of the equation
${d\phi_s(p)\over ds}=X\circ\phi_s(p)$ with initial condition
$\phi_0(p)=p)$ is defined for all $s\in
(t_-(p),t_+(p))$.
{}From continuous dependence of solutions of ODE's upon parameters it follows
that for every $\delta>0$ there exists a neighborhood ${\cal
O}_{p,\delta}$ of $p$
such that
for all $q\in  {\cal O}_{p,\delta}$ we have $t_+(q)\geq t_+(p)-\delta$ and
$t_-(q)\leq t_-(p)+\delta$. In other words, $t_+$ is a lower semi-continuous
function and $t_-$ is an upper semi-continuous function. We have the following:

\begin{Lemma}
\label{L1.1}  Let $p\in  I^+(\veps)$, $ q \in  J^-(p) \cap I^+
(\veps)$, suppose that $t_+(p)\geq \tau_0$. 
If $t_+(q) < \tau_0$, 
then there
exists $s\in [0,t_+(q))$ such that $\phi_s(q)\in \veps$.
\end{Lemma}

{\bf Proof:} Let $\gamma(s)$ be any  future directed causal curve with
$\gamma(0)=q$, $\gamma(1)=p$. Suppose that $t_+(p)\geq\tau_0>t_+(q)$, let
$(s_-,1]$ be the largest  interval such that $t_+(\gamma(s))>t_+(q)$
for all $s\in (s_-,1]$. By lower semi-continuity of $t_+$ we have
$(s_-,1]\neq\phie$.  Consider the one-parameter family of causal paths
$$
[0,t_+(q)]\times(s_-,1]\ni (\tau,s)\to\tilde\gamma_\tau(s)=\phi_\tau
(\gamma(s))   .
$$
Suppose that for all $s\in [0,t_+(q))$ we have $\phi_s(q)\not\in
\Sigma$. Global
hyperbolicity of $M$ implies that for all $s\in [0,t_+(p))$ we have
$\phi_s(q)\in  I^+(\Sigma)$, consequently for any $r\in
I^+(\phi_s(q))$ it also
holds that $r\in  I^+(\veps)$; hence $\tilde\gamma_\tau(s)\in  J^+(\Sigma)$ for
all $\tau, s\in [0, t_+(q)]\times(s_-,1]$. As $\Sigma$ is a Cauchy
surface, for
each $\tau$ the curve $\tilde\gamma_\tau$ must be past-extendible. Let thus
$\hat\gamma_\tau(s)$ be any past extension of $\tilde\gamma_\tau$,
for $\tau\in [0, t_+(q)]$ define
$\psi_\tau=\hat\gamma_\tau(s_-)$. It is elementary to show that
$\psi_\tau=\phi_\tau(\gamma(s_-))$, so that $t_+(\gamma(s_-))>t_+(q)$. This,
however, contradicts the definition of $s_-$, and the result follows.
\hfill $\Box$

{\bf Proof of Theorem \ref{T1}:} Suppose  there exists $s_0\in
[-\eps,\eps]$ such
that $\phi_{s_0}(\veps)$ is {\it not} achronal. Let $\Gamma:[0,1]\to M$ be a
timelike curve such that $\Gamma(0),\Gamma(1)\in \phi_{s_0}(\veps)$. Changing
$X$ to $-X$ if necessary we may assume $s_0<0$; changing time orientation if
necessary we may suppose that $\Gamma(1) \in  J^+(\veps)$. We have
$t_+|_\Sigma\geq\eps$, hence
$t_+|_{\phi_{s_0}(\Sigma)}=(t_++|s_0|)|_\veps\geq \eps$.
%

Let $q \in I^-(\Gamma(1))\cap J^+(\veps)$. By Lemma \ref{L1.1} either
$t_+(q)\geq \eps$, or
there exists $s\in [0, t_+(q))$ such that $\phi_s(q)\in \veps$. In
that last case
we have $t_+(q)-s=t_+(\phi_s(q))\geq\eps$, hence $t_+(q)\geq\eps$, and in
either case we obtain $t_+(q)\geq\eps$.
It follows that
\be
t_+|_{\Gamma\cap J^+(\veps)}\geq\eps   .
\label{(AAA.0)}
\ee
If $\Gamma(0)\in  J^+(\veps)$ we thus obtain
\be
t_+|_\Gamma \geq\eps .
 \label{(AAA.1)}
\ee
Consider the case $\Gamma(0)\in  J^-(\veps)$. We have $t_+(\Gamma(0))\geq\eps$
and by an argument similar to the one above (using the time-dual version of
Lemma \ref{L1.1}) we obtain 
$$
t_+|_{\Gamma\cap J^-(\veps)}\geq\eps ,
$$
and by global hyperbolicity we can again conclude that (\ref{(AAA.1)})
holds. Eq.\ (\ref{(AAA.1)})
shows that $\phi_{-s_0}(\Gamma)$ is a 
timelike curve satisfying
$\phi_{-s_0}(\Gamma(0)),\phi_{-s_0}(\Gamma(1))\in \veps$.
This, however, contradicts achronality of $\veps$. We therefore conclude that
for all $s\in [-\epsilon,\epsilon] $ the hypersurfaces $\phi_s(\veps)$
are achronal.
Theorem~\ref{T1} follows now from Theorem~\ref{T1.0}.\hfill $\Box$


\section{Proof of Corollary {\protect\ref{C1}}}
\label{ProofsC1}
Before passing to the proof of Corollary \ref{C1}, it seems
appropriate to present some definitions.

\begin{Definition}
\label{D1}
We shall say that an initial data set $(\veps,\gamma,K)$ for vacuum Einstein
equations is asymptotically flat if $(\veps,\gamma)$ is a complete
Riemannian manifold (without boundary), with $\Sigma$ of the form
\begin{equation}
\label{topological}
\veps=\veps_{\rm int} \bigcup^I_{i=1}\veps_i,
\ee
for some $I<\infty$. Here we assume that $\veps_{\rm int}$ is compact, and each
of
the ends $\veps_i$ is diffeomorphic to $\R^3\setminus B(R_i)$ for some
$R_i>0$, with
$B(R_i)$ --- coordinate ball of radius $R_i$. In each of the ends $\veps_i$ the
metric is assumed to satisfy the hypotheses\footnote{The
differentiability threshold of Theorem~6.1 of \cite{ChOM} can actually be
weakened to $s\geq3$. Similarly, the differentiability threshold in Theorem~6.2
of \cite{ChOM} can be weakened to $s\geq4$, and probably also to
$s\geq3$. }
of the boost theorem, Theorem~6.1 of \cite{ChOM}.
\end{Definition}

The hypotheses of Theorem~6.1 of \cite{ChOM} will hold if {\it e.g.\/}
there exists $\a>0$ such that in each of the ends $\veps_i$ we have
$$
0\leq k\leq 4 
\quad
|\part_{i_1}\dots\part_{i_k}(\gamma_{ij}-\delta_{ij})| \leq
Cr^{-\a-k} , 
$$
$$
0\leq k\leq 3 
 \quad |\part_{i_1}\dots\part_{i_k} K_{ij}|\leq Cr^{-\a-k-1},
$$
for some constant $C$.

To motivate the next definition, consider a space--time with some
number of asymptotically flat ends, and with a black hole region. In
such a case there might be a Killing vector field defined in, say, the
domain of outer communication of the asymptotically flat ends. It
could. however, occur, that there is no Killing vector field defined
on the whole space--time --- a famous example of such a space--time
has been considered by Brill \cite{Brill}, yielding a space--time in
which no asymptotically flat maximal surfaces exist. Alternatively,
there might be a Killing vector field defined everywhere, however, there
might be some non-asymptotically flat ends in $M$. [As an example.
consider a spacelike surface in the Schwarzschild--Kruskal--Szekeres
space--time in which one end is asymptotically flat, and the second is
``asymptotically hyperboloidal''.] In such cases one would still like
to claim that the orbits of $X$ are complete at least in the exterior region.
We shall see that this is indeed the case, under some conditions which
we spell out below:

\begin{Definition}
\label{D2}
Consider a stably causal Lorentzian manifold $(M,g)$ with an
achronal  spacelike
surface $\hat\Sigma$.  Let $\Sigma \subset \hat\Sigma$ be a connected
submanifold
of $\hat \Sigma$ with smooth compact boundary $\partial\Sigma$, and
let $(\gamma,K)$ be the Cauchy data induced by $g$ on $\Sigma$.
Suppose finally that there exists a Killing vector field $X$ defined on
$D(\Sigma)$. We
shall say that $(\Sigma,\gamma,K)$ are
Cauchy data for an asymptotically flat
exterior region in a (non--degenerate) black--hole space--time
if the following hold:
\begin{enumerate}
\item
The closure $\bar \Sigma\equiv \Sigma\cup\partial\Sigma$ of $\Sigma$
is of the form (\ref{topological}), with $ \veps_{\rm int}$ and
$\veps_i$ satisfying the requirements of Definition \ref{D1}.

\item
\,
[From eq.\ (\ref{Xequation}) below it follows that $X$ can be extended
by continuity to $\overline{ D(\Sigma)}$.] We shall require that $X$ be
tangent to $\partial \Sigma$.
\end{enumerate}
\end{Definition}

An example of the behaviour described in Definition \ref{D2} can be
observed in the Schwarzschild--Kruskal--Szekeres space--time $M$, when
$\hat \Sigma$ is taken as a standard $t=0$ surface, $\Sigma$ is the
part of $\hat\Sigma$ which lies in  one asymptotic end of $M$, and
$\partial \Sigma$ is the set of points where  the usual Killing
vector $X$ (which coincides with $\partial/\partial t$ in the
asymptotic regions) vanishes. Such $\partial\Sigma$'s are usually
called ``the bifurcation surface of a bifurcate Killing horizon''. An
example in which $X$ does not vanish on $\partial\Sigma$ is given by
the Kerr space--time, when $X$ is taken to coincide with
$\partial/\partial t$ in the asymptotic region, and $\partial \Sigma $
is the intersection of the black hole and of the white hole with
respect to the asymptotic end under consideration.

The notion of {\em non--degeneracy} referred to in definition \ref{D2}
above is related to the non--vanishing of the surface gravity  of the
horizon:  Indeed, it follows from \cite{WR} that 
in situations of interest the behaviour described in Definition
\ref{D2} can only occur if the
surface gravity of the horizon is constant  on the horizon, and does
not vanish. 

 With the above definitions in mind, we can now prove
Corollary~\ref{C1}:

{\bf Proof of Corollary \ref{C1}:}
Suppose first that $\veps$ is compact. We have
$$
t_+|_\veps \geq\eps
$$
for some $\eps>0$, because a lower semi-continuous function attains its infimum
on a compact set ({\it cf.\ e.g.\/} \cite{Struwe}), and the result follows
from Theorem~\ref{T1}.
[Here we could also use Theorem~\ref{T1.0}: the hypersurfaces $\phi_s(\veps),
s\in [-\eps,\eps]$, are compact and spacelike and hence achronal by
\cite{BILY}.]

Consider next the case $(\veps,\gamma,K)$ -- asymptotically flat. Let
$(M,g)$
be the maximal globally hyperbolic development of $(\veps,\gamma
  ,K)$. A
straightforward extension of the boost-theorem \cite{ChOM} using domain of
dependence arguments shows that $M$ contains a subset of the form
\be
M_1=([-\delta,\delta]\times\veps_{\rm int})\bigcup^I_{i=1}\Omega_i,
\label{(AAA.2)}
\ee
with some $\delta>0$, where each of the $\Om_i$'s is a boost-type domain:
$$
\Omega_i=\{(t,\vec{x})\in  \R^4:|\vec x|\geq R_i,|t|\leq\delta+\theta
(r-R_i)\},
$$
with some $\theta>0$. Let $X$ be a Killing vector field on $M$. As is well
known, $X$ satisfies the equations
\be
\label{Xequation}
\nabla_\mu\nabla_\nu X_\a=R_{\la\mu\nu\a}X^\lambda.
\ee
Under the hypotheses of Theorem~6.1 of
\cite{ChOM}, a  simple analysis of 
(\ref{Xequation}) shows that in each $\Omega_i$ there exists $\alpha>0$ and a
constant (perhaps vanishing) matrix
${\Lambda^\mu}_\nu={\Lambda^\mu}_\nu(i)$ such that 
\be
0\leq j\leq2\quad \part_{i_1}\dots\part_{i_j} [X^\mu-{\Lambda^\mu}_\nu x^\nu]=O
(r^{1-\a-j}) \label{(AAA.3)}
\ee
{}From equations (\ref{(AAA.2)}) and (\ref{(AAA.3)}) one easily shows
that there exists
$\epsilon>0$ such that for
all $p\in \veps$ the orbit $\phi_s(p)$ of $X$ through $p$ remains in $M_1$ for
$|s|\leq\eps$. This shows that in the asymptotically flat case the hypotheses
of Theorem~\ref{T1} are satisfied as well, and the second part of
Corollary~\ref{C1} follows.

Consider finally point 3 of Corollary~\ref{C1}. Let $\hX$ be any
vector field (not necessarily Killing) defined in a neighbourhood
$\co$ of
$\partial \Sigma$  such that $\hX|_{\co\cap D(\Sigma)}=X$.

[Because $D(\Sigma)$ is
{\em not} a smooth manifold, a little work is needed to show that an
extension $\hX$ of $X$ exists. A possible construction goes as
follows:  Define $\psi^\mu=X^\mu|_\Sigma$,
$\chi^\mu=n^\alpha\nabla_\alpha X^\mu|_\Sigma$, where $n^\alpha$ is
the field of unit normals to $\Sigma$. Because $\partial \Sigma$ is smooth in
$\hat \Sigma$, there exists smooth extensions $\hat \chi^\mu$ and
$\hat\psi^\mu$ of $\chi^\mu $ and of $\psi^\mu$ from $\Sigma$ to $\hat
\Sigma$. 
On $D(\hat
\Sigma)$ 
let $\hat X 
$ 
be the unique solution of the problem
\be
\label{problem}
\left.
\matrix{\Box \hat X^\mu = -{R^\mu}_\alpha \hat X^\alpha 
\cr
\hat X^\mu\Big|_{\hat \Sigma} = \psi^\mu, \quad
\hat n^\alpha\nabla_\alpha \hat X^\mu\Big|_{\hat \Sigma} = \chi^\mu 
\cr }
\right\}
\ee
where $\hat n^\alpha$ is the field of unit normals to $\hat \Sigma$ and
${R^\mu}_\alpha$ is the Ricci tensor of $g$. We have $\hat
X|_{D(\Sigma)}=X$
by uniqueness of solutions of (\ref{problem}).]

Returning to the main argument, without loss of generality we may
assume that the neighbourhood $\co$ of $\partial \Sigma$ is covered by
normal geodesic coordinates based on $\partial \Sigma$:
$$
\co = \{(q,t,x): q\in\partial\Sigma, (t,x)\in B(\epsilon)\subset \R^2
\},
$$
for some $\epsilon >0$, where $B(\epsilon)$ is a coordinate ball of
radius $\epsilon$. We have $\partial \Sigma\cap \co =
\{(q,t,x):t=x=0\}$, and we can also assume that $\overline \co$ is a
compact subset of $M$. For $p\in\co$ and $s\in(\hat t_-(p),\hat
t_+(p))$ let $\hat \phi _s (p)$ be the orbit of $\hat X$ through $p$.
There exists $\epsilon >0$ such that $\hat
t_+(p) |_\co \ge \epsilon$,  $\hat
t_-(p) |_\co \le -\epsilon.$ Consider\footnote{The argument that
follows is essentially due to R.\ Wald.} $p\in \co\cap D(\Sigma)$,
thus $p=(q,t,x)$, with  $q\in\partial \Sigma$, $(t,x)\in B(\epsilon)$;
changing $x$ to $-x$ if necessary we also have $|t|< x$. By
construction of the coordinates $(q,t,x)$ the straight lines $q=q_0$,
$t=\alpha s$, $x = \beta s$, $\alpha,\beta\in\R$, are affinely
parametrized geodesics. Now
for $|s|\le \epsilon$ 
$\hat \phi_s: \co\cap D(\Sigma)\rightarrow M$ are isometries, hence in
$\co\cap D(\Sigma)$ the maps $\hat \phi_s$ carry geodesics
into geodesics and preserve affine parametrization. It follows  that
the $\hat
\phi_s$'s must be of the form
$$
\co\cap D(\Sigma) \ni (q,x^\mu) \rightarrow \phi_s(q,x^\mu)=\hat
\phi_s(q,x^\mu)=(\psi_s(q), {\Lambda(s,q)^\mu}_\nu x^\nu),
$$
for some map $\psi_s:\partial \Sigma\rightarrow \partial \Sigma$,
where we have set $x^\mu=(t,x)$, and where $\Lambda(s,q)$ is a Lorentz
boost. Consequently, we can find $0<\delta\le\epsilon$ and a
conditionally compact neighbourhood
$\cu$ of $\partial \Sigma$, $\cu\subset \co$, such that for all $p\in
{ \cu\cap\Sigma}$ and for $s\in[-\delta,\delta]$ we shall
have $\phi_s(p)\in D(\Sigma)$. The result follows now from the
arguments of the proof of parts 1 and 2 of this Corollary.
\hfill $\Box$

\end{document}